\pdfoutput=1

\documentclass[reqno,11pt]{article}
\usepackage{jheppub}
\usepackage{epsfig}
\usepackage{amssymb}
\usepackage{amsmath}
\usepackage{hyperref}
\usepackage{dsfont}
\usepackage{slashed}
\usepackage[dvipsnames]{xcolor}

\newcommand{\be}{\begin{eqnarray}}
\newcommand{\ee}{\end{eqnarray}}
\newcommand{\soft}{\tau}

\title{Are Soft Theorems Renormalized?}
\author[a]{Freddy Cachazo}
\author[a,b]{and Ellis Ye Yuan}
\affiliation[a]{Perimeter Institute for Theoretical Physics,\\31 Caroline Street North, Waterloo, ON N2L 2Y5, Canada}
\affiliation[b]{Department of Physics \& Astronomy, University of Waterloo,\\200 University Avenue West, Waterloo, ON N2L 3G1, Canada}
\emailAdd{fcachazo@pitp.ca}
\emailAdd{yyuan@pitp.ca}

\abstract{We show that the distributional nature of soft theorems requires the soft limit expansion to take priority over the regulator expansion of Feynman loop integrals. We start the study of soft graviton theorems at loop level from this perspective by considering a five-particle one-loop amplitude in ${\cal N}=8$ supergravity. Surprisingly, we find that a soft theorem recently introduced by one of the authors and Strominger is not renormalized in this case. Computations are done in $4-2\epsilon$ dimensions and for terms of order $\epsilon^{-2}$, $\epsilon^{-1}$ and $\epsilon^{0}$.}

\begin{document}
\maketitle


\section{Introduction}\label{sect:intro}


Recently one of the authors and Strominger proposed a new soft theorem for gravity amplitudes \cite{Cachazo:2014fwa}, which extends the soft theorem by Weinberg up to sub-sub-leading terms \cite{Weinberg:1964ew,Weinberg:1965nx}. The idea stemmed from the study of the BMS group \cite{Strominger:2013un,Strominger:2013jfa,He:2014laa}, which conjectures a new infinite dimensional symmetry of the gravitational S-matrix.
The full proposal reads
\begin{equation}\label{newtheorem}
\mathcal{M}_{n+1}(\soft)=\left(\frac{1}{\soft^3}S^{(0)}+\frac{1}{\soft^2}S^{(1)}+\frac{1}{\soft}S^{(2)}\right)\,\mathcal{M}_n+
\mathcal{O}(\soft^0),
\end{equation}
where the particle taken to be soft is $s=n+1$ and $\lambda_s(\soft )\equiv \soft\lambda_s$, with $\soft$ the parameter that controls the soft limit\footnote{In this paper we use the notation $\soft$ for the soft limit parameter, which differs from that in \cite{Cachazo:2014fwa}, in order not to confuse it with the dimensional regularization parameter $\epsilon$ at loop level.} (For earlier work on this topic, see \cite{Low:1958sn,Burnett:1967km,Gross:1968in,Jackiw:1968zza,White:2011yy}). We restrict our discussion to four dimensions where the spinor-helicity formalism allows us to consider a holomorphic soft limit when the soft particle is a positive helicity graviton.

One of the most important features of (\ref{newtheorem}) is that on both sides of the equation $\mathcal{M}_{m}$ denotes amplitudes containing momentum conserving delta functions. We make this explicit by introducing the notation $\mathcal{M}_m = M_m\,\delta^4(\sum_{a=1}^m k_a)$. In other words, (\ref{newtheorem}) is a relation among distributions. This observation is crucial for the line of thought pursued in this work.

The explicit expressions of the soft operators $S^{(0)},S^{(1)},S^{(2)}$ in spinor-helicity form are
\begin{align}
S^{(0)}&=\sum_{a=1}^n\frac{[s,a]}{\langle s,a\rangle}\frac{\langle x,a\rangle\langle y,a\rangle}{\langle x,s\rangle\langle y,s\rangle},\\
S^{(1)}&=\frac{1}{2}\sum_{a=1}^n\frac{[s,a]}{\langle s,a\rangle}\left(\frac{\langle x,a\rangle}{\langle x,s\rangle}+\frac{\langle y,a\rangle}{\langle y,s\rangle}\right)\tilde\lambda_s^{\dot\alpha}\frac{\partial}{\partial\tilde\lambda_a^{\dot\alpha}},\\
S^{(2)}&=\frac{1}{2}\sum_{a=1}^n\frac{[s,a]}{\langle s,a\rangle}\,\tilde\lambda_s^{\dot\alpha}\tilde\lambda_s^{\dot\beta}\frac{\partial^2}{\partial\tilde\lambda_a^{\dot\alpha}\,\partial\tilde\lambda_a^{\dot\beta}},
\end{align}
with $x$ and $y$ arbitrary holomorphic reference spinors.


Weinberg's soft theorem in gravity is known not to be renormalized at loop level \cite{Bern:1998sv}. It is then natural to investigate the behavior of the sub-leading corrections at loop level as well.

Some results have already appeared in the literature. It has been shown up to twelve particles that (\ref{newtheorem}) is valid in the all-plus sector \cite{He:2014bga} at one loop. Also, the study of amplitudes in the one-minus sector \cite{He:2014bga} and of explicit IR divergent terms at one loop \cite{Bern:2014oka} imply that (\ref{newtheorem}) must receive corrections. All these checks have been performed on integrated amplitudes which are expanded in $\epsilon$ before taking the soft limit.

In this paper we show that at loop level the distributional nature of (\ref{newtheorem}) forces the soft limit expansion to be performed before any expansion in the regulator parameter. Moreover, the soft limit must take priority over any expansions or limits in the definition of the loop amplitudes. Here we work in dimensional regularization, i.e., $D=4-2\epsilon$, and therefore all expansions in $\soft$ must be performed before Feynman integrals are expanded in $\epsilon$. This order of limits is the opposite to the one conventionally adopted in the literature and can lead to dramatically different results.

In this note we present evidence that when the order imposed by the distributional nature of the soft theorems is used it is possible for them to hold exactly at loop level. This does not necessarily mean that soft theorems impose more constraints on amplitudes than the standard expansions around soft limits as the corrections that appear in the latter have a simple structure \cite{Bern:2014oka}.

The main computation in this paper is the study of the soft limit expansion of a five-particle amplitude in ${\cal N}=8$ supergravity which is IR divergent and hence it has to be defined in $4-2\epsilon$ dimensions. We start with a formula valid to all order in $\epsilon$ for ${\cal M}_5^{\rm 1-loop}$ and perform the $\soft$ expansion at the level of the integrand. The reason for this is that integrals that normally contribute to order $\epsilon$ can contribute to lower orders when the soft limit is done first.

The resulting object is then integrated using Mellin-Barnes techniques as an expansion in $\epsilon$. We also compute the action of $S^{(0)}$ and $S^{(1)}$ on a four-particle amplitude ${\cal M}_4^{\rm 1-loop}$. Surprisingly, it turns out that
\be
{\cal M}_5^{\rm 1-loop}(\soft ) = \left(\frac{1}{\soft^3}S^{(0)}+\frac{1}{\soft^2}S^{(1)}\right)\,\mathcal{M}_4^{\rm 1-loop}+
\mathcal{O}(\soft^{-1}) .
\ee
More explicitly, we carried out the expansion in $\soft$ to order $\soft^{-2}$ and in $\epsilon$ to order $\epsilon^{0}$, and the equation above holds at each order. We leave the computation at higher orders for future work.

We end by discussing how the distributional nature of the soft theorems for gravity can be extended to gauge theory, as well as to other relations among S-matrix elements that follow from unitarity such as factorization.

This paper is organized as follows. In Section 2, we discuss how the distributional nature of the amplitudes entering in the soft theorems uniquely singles out the way soft limits and regulator expansions must be performed. In Section 3 we present the main computation of the paper: the computation of the soft limit expansion of the five-particle one-loop amplitudes in ${\cal N}=8$ supergravity and its comparison with the soft theorem (\ref{newtheorem}). In Section 4, we discuss the same computation as in Section 3 but done with a formula for the five-particle amplitude solely in terms of scalar boxes. Such a formula is known to be valid up to order $\epsilon$ terms. The result is that such a formula does not reproduce (\ref{newtheorem}) and the discrepancy is accounted for by a pentagon in $6-2\epsilon$ dimensions as expected from the results in Section 3. In section 5, we comment on the distinction between the soft theorems as discussed in this work and the soft expansions which are standard in the literature. Discussions and future directions are presented in Section 6. The appendix contains some of the details of the computation of scalar integrals.

\section{Precise Definition of Soft Theorems and Regulator Expansion}\label{sect:precisedefinition}

The gravitational soft theorems studied in \cite{Cachazo:2014fwa}, i.e.,
\begin{equation}
\mathcal{M}_{n+1}(\soft)=\left(\frac{1}{\soft^3}S^{(0)}+\frac{1}{\soft^2}S^{(1)}+\frac{1}{\soft}S^{(2)}\right)\,\mathcal{M}_n+
\mathcal{O}(\soft^0)
\end{equation}
are meant to be relations among distributions. Any relation among distributions has to be valid for any test function\footnote{Test functions are defined to be real functions with continuous derivatives of all orders and with bounded support.}. In other words, two distributions are equal if they give the same result when integrated against any arbitrary test function (see e.g., \cite{gelfand}). This is specially important when dealing with distributions with singular support such as scattering amplitudes.

In order to make the soft theorem (\ref{newtheorem}) mathematically precise it has to be decomposed into three theorems. The reason for the decomposition will become clear shortly.

The leading order, i.e., Weinberg's theorem, becomes the following
\begin{equation}\label{theorem0}
\underset{\soft\to 0}{\lim} \left( \soft^3 {\cal M}_{n+1}(\soft\lambda_s)\right) = S^{(0)}\mathcal{M}_n.
\end{equation}
The sub-leading order theorem is
\begin{equation}\label{theorem1}
\underset{\soft\to 0}{\lim} \left( \soft^2 {\cal M}_{n+1}(\soft\lambda_s ) - \frac{1}{\soft} S^{(0)}\mathcal{M}_n\right) = S^{(1)}\mathcal{M}_n.
\end{equation}
And finally, the sub-sub-leading order theorem is
\begin{equation}\label{theorem2}
\underset{\soft\to 0}{\lim} \left( \soft {\cal M}_{n+1}(\soft\lambda_s ) - \frac{1}{\soft^2} S^{(0)}\mathcal{M}_n- \frac{1}{\soft^2}S^{(1)}\mathcal{M}_n\right)  = S^{(2)}\mathcal{M}_n.
\end{equation}
This new form of the soft theorems seems to be completely equivalent to the previous one (\ref{newtheorem}). Indeed, both definitions are equivalent for generic test functions. To understand the difference, note that (\ref{newtheorem}) implicitly assumes that $\soft$ is small but finite as one does with any expansion. In the new form, (\ref{theorem0}), (\ref{theorem1}) and (\ref{theorem2}), the relations are stated in the strict limit $\soft\to 0$. While both formulations agree for generic test functions, they could differ in cases where they are integrated against test functions with support smaller than the small but finite $\soft$ used in (\ref{newtheorem}). Now it is clear why the theorems, as distributions, must be defined in the strict limit; only in the limit the theorem is valid for all test functions.

The natural question at this point is whether anything is to be gained by these mathematically precise definitions. As explained in the introduction, the distributional nature of the soft theorems will single out a unique order of limits when dealing with loop amplitudes defined with a regulator.

\subsection{Regulator Expansion and Soft Theorems}

In any interacting quantum field theory in four dimensions one has to introduce a regulator for the definition of its S-matrix due to the presence of divergences. To be definite, let us consider dimensional regularization in the scheme where all external data are kept four dimensional \cite{Bern:1991aq}.

Scattering amplitudes are expanded around $\epsilon=0$ and in general the expansions have poles in $\epsilon$, reflecting the presence of divergences. This means that the regulator has to be kept finite at all times until physical observables are computed.

It is now clear that soft theorems, as relations among S-matrix elements of momentum eigenstates, have to be defined and understood among regulated amplitudes, i.e., with small but finite $\epsilon$. As a consequence, when studying soft theorems of the form \eqref{theorem0}, \eqref{theorem1} and \eqref{theorem2}, one has to keep $\epsilon$ finite while the limit $\soft\to 0$ is taken. For example, the first two theorems read
\begin{equation}
\underset{\soft\to 0}{\lim} \left( \soft^3 {\cal M}_{n+1}^{D=4-2\epsilon}(\soft\lambda_s)\right) = S^{(0)}\mathcal{M}_n^{D=4-2\epsilon},
\end{equation}
and
\begin{equation}
\underset{\soft\to 0}{\lim} \left( \soft^2 {\cal M}_{n+1}^{D=4-2\epsilon}(\soft\lambda_s ) - \frac{1}{\soft} S^{(0)}\mathcal{M}_n^{D=4-2\epsilon}\right) = S^{(1)}\mathcal{M}_n^{D=4-2\epsilon}.
\end{equation}
In practice this implies that in both theorems $\soft$ is always taken to be smaller than $\epsilon$ and therefore any expansion of Feynman loop integrals should first be done in $\soft$ and then in $\epsilon$.

The fact that these two expansions do not commute is well known \cite{Bern:1995ix}. In order to illustrate this point let us discuss two examples that will also be useful in the rest of the paper.

\subsection{Examples}\label{sect:twoexamples}

The first example is very standard. Consider a scalar box integral with momenta $k_1$, $k_2$, $k_3$ and $k_4+k_5$ at the corners. As usual all momenta are taken to be those of on-shell particles, i.e., $k_a^2=0$. The integral is defined as
\be\label{int5}
{\cal I}^{123(45)}_4 =\int\frac{d^{4-2\epsilon}L}{N}\frac{1}{(L+k_1+k_2)^2\,(L+k_2)^2\,L^2\,(L-k_3)^2}\delta^4(\sum_{a=1}^5k_a),
\ee
where we inserted the normalization factor $N=2i\pi^{\frac{D}{2}}e^{-\gamma_E\epsilon}$ in order to make the $\epsilon$-expansion simpler in appearance. The subscript $4$ indicates that the integral is a box, i.e., it has four propagators. The notation $123(45)$ indicates the momenta flowing out of each vertex. For example, at one of the vertices one has $k_4+k_5$ flowing out of the vertex. This integral is UV finite but IR divergent. An expansion around $\epsilon=0$ reveals that
\begin{equation*}
I^{123(45)}_4 = \frac{1}{st}\left( \frac{1}{\epsilon^2}(-s)^{-\epsilon} + \frac{1}{\epsilon^2}(-t)^{-\epsilon} -\frac{1}{\epsilon^2}(-2k_4\cdot k_5)^{-\epsilon} \right) + {\cal O}
(\epsilon^0),
\end{equation*}
with $s=(k_1+k_2)^2$ and $t=(k_2+k_3)^2$ the standard kinematic invariants\footnote{Here we have set a mass scale generated in dimensional regularization to one. This is what makes the arguments of $(-x)^{-\epsilon}$ dimensionless.}. Here we use again the convention that ${\cal I}_4$ denotes the integral with a momentum conserving delta function while $I_4$ is the stripped one. Further expanding in $\epsilon$ one has
\begin{equation*}
I^{123(45)}_4 = \frac{1}{st}\left( \frac{1}{\epsilon^2} +\frac{1}{\epsilon}\left( -\log(-s)-\log(-t)+\log(-2k_4\cdot k_5)\right)\right) + {\cal O}(\epsilon^0 ).
\end{equation*}
In the soft limit $k_5\to 0$ we see a logarithmic divergent term.

Consider the expansions in the opposite order. First take the strict soft limit $k_5 \to 0$ in (\ref{int5}). This gives
\be
{\cal I}^{1234}_4\equiv \underset{k_5\to 0}{\lim}{\cal I}^{123(45)}_4 =\int\frac{d^{4-2\epsilon}L}{N}\frac{1}{(L+k_1+k_2)^2\,(L+k_2)^2\,L^2\,(L-k_3)^2}\delta^4(\sum_{a=1}^4k_a).\nonumber
\ee
This new integral is a scalar box with momenta $\{k_1,k_2,k_3,k_4\}$ at its corners. Now we can expand around $\epsilon =0$ to get
\be
I^{1234}_4 = \frac{1}{st}\left( \frac{1}{\epsilon^2}(-s)^{-\epsilon} + \frac{1}{\epsilon^2}(-t)^{-\epsilon} \right) + {\cal O}
(\epsilon^0) = \frac{1}{st}\left( \frac{2}{\epsilon^2}+\frac{1}{\epsilon}\left( -\log(-s)-\log(-t)\right)\right)+{\cal O}
(\epsilon^0).\nonumber
\ee
For fixed $\epsilon$ this new expression is completely finite and therefore it shows that expansions in $\soft$ and in $\epsilon$ need not commute.

Our second example is less well-known but more striking. Consider a scalar pentagon integral with massless momenta flowing out of
all of its corners. In our example the external momenta are four dimensional while the loop integration is defined in $6-2\epsilon$ dimensions. The choice of dimension seems puzzling at first but reduction techniques of Feynman integrals in $4-2\epsilon$ dimensions can lead to integrals with shifted dimensions \cite{Bern:1998sv}. This is indeed the reason this particular example will be useful in Section \ref{sect:correctformula}.

The integral under consideration is
\be
{\cal I}_5^{12345} =\!\int\frac{d^{6-2\epsilon}L}{N}\frac{1}{(L+k_2+k_3)^2\,(L+k_3)^2\,L^2\,(L-k_4)^2\,(L-k_4-k_5)^2}\delta^4(\sum_{a=1}^5k_a).\quad
\ee
It is easy to show that this integral is free of both UV and IR divergences. Therefore the integral at $\epsilon=0$ can be performed to produce a well-defined function of kinematic invariants and in particular of $k_5\cdot k_a$ (see \cite{Kniehl:2010aj}). One could now take the soft limit $k_5\to 0$, but we will not compute it as it is not needed for our purposes.

Consider now the opposite order, i.e., first take the soft limit $k_5\to 0$,
\begin{equation*}
\underset{k_5\to 0}{\lim}{\cal I}_5^{12345} =\int\frac{d^{6-2\epsilon}L}{N}\frac{1}{(L+k_2+k_3)^2\,(L+k_3)^2\,L^2\,((L-k_4)^2)^2}\delta^4(\sum_{a=1}^4k_a).
\end{equation*}
Using Mellin-Barnes techniques for the computation of loop Feynman integrals it is possible to show that
\begin{equation*}
\int\frac{d^{6-2\epsilon}L}{N}\frac{1}{(L+k_2+k_3)^2\,(L+k_3)^2\,L^2\,((L-k_4)^2)^2} = -\frac{1}{2st}\epsilon^{-2}+\frac{\log (-t)}{2st}\epsilon^{-1}+ {\cal O}(\epsilon^0 ),
\end{equation*}
where again $s=(k_1+k_2)^2$ and $t=(k_2+k_3)^2$.

These two examples show that the order of expansions can have dramatic consequences in the result.

\section{Application to a Five-Point One-Loop Amplitude in ${\cal N}=8$ Supergravity}

In the rest of this paper we consider a five-point amplitude of gravitons in ${\cal N}=8$ supergravity. This amplitude is chosen because it is the simplest non-trivial example that illustrates all the subtleties of the limits. The amplitude is UV finite but IR divergent.

Let the one-loop $m$-point MHV amplitude in ${\cal N}=8$ supergravity be given by
\be
{\cal M}_m^{D=4-2\epsilon} =M_m^{D=4-2\epsilon}\,\delta^{0|16}(\tilde\eta_1\lambda_1+\cdots + \tilde\eta_m\lambda_m )\,\delta^4(k_1+\cdots +k_m).
\ee
In this work we will focus on the $5$-point and $4$-point amplitudes and consider soft limits where $k_5\to 0$. Since we are interested in using a holomorphic soft limit, it is important to take particle $5$ to be a graviton of positive helicity. Having positive helicity means that we should set $\tilde\eta_5 =0$ and therefore the supersymmetric delta function becomes irrelevant in our discussions and will be dropped from this point on.

The main result of this paper is to show that
\begin{equation}\label{result0}
\underset{\soft\to 0}{\lim} \left( \soft^3 {\cal M}_{5}^{D=4-2\epsilon}(\soft\lambda_s)\right) = S^{(0)}\mathcal{M}_4^{D=4-2\epsilon} ,
\end{equation}
and
\begin{equation}\label{result1}
\underset{\soft\to 0}{\lim} \left( \soft^2 {\cal M}_{5}^{D=4-2\epsilon}(\soft\lambda_s ) - \frac{1}{\soft} S^{(0)}\mathcal{M}_4^{D=4-2\epsilon }\right) = S^{(1)}\mathcal{M}_4^{D=4-2\epsilon}
\end{equation}
hold at least up to order $\epsilon$. We leave the computation of higher order terms in $\epsilon$ as well as at the sub-sub-leading order in $\soft$, i.e., (\ref{theorem2}) for future work.

In these formulas we use a representation for $M_5$ valid to all orders in $\epsilon$. Fortunately, such a formula is explicitly known and is relatively simple \cite{Cachazo:2008vp,Carrasco:2011mn,Yuan:2012rg},
\begin{equation}\label{eq:5ptformula}
M_5=\left(\frac{[1,2][2,3][3,4][4,5][5,1]}{(1,2,3,4)}\right)^2I_5^{12345}+\frac{[4,5]^3}{\langle4,5\rangle}\left(\frac{[1,2][2,3][3,1]}{(1,2,3,4)}\right)^2I_4^{123(45)}+
\text{Permutations},
\end{equation}
where $I_5^{12345}$ denotes the usual scalar pentagon integral with the ordering $(12345)$, and $I_4^{123(45)}$ the usual scalar box integral with one massive corner attached by particles $4$ and $5$, and all integrals are evaluated in $4-2\epsilon$ dimensions. Explicitly,
\begin{align}
I_5^{12345}&=\int\frac{d^{4-2\epsilon}L}{N}\,\frac{1}{(L+k_2+k_3)^2\,(L+k_3)^2\,L^2\,(L-k_4)^2\,(L-k_4-k_5)^2},\\
I_4^{123(45)}&=\int\frac{d^{4-2\epsilon}L}{N}\,\frac{1}{(L+k_1+k_2)^2\,(L+k_2)^2\,L^2\,(L-k_3)^2},
\end{align}
where the normalization factor $N=2i\pi^{\frac{D}{2}}e^{-\gamma_E\epsilon}$. The permutations in \eqref{eq:5ptformula} are performed over all inequivalent configurations of the labeled pentagons and boxes (hence altogether $12$ pentagons and $30$ boxes), and the totally anti-symmetric combination\footnote{This is usually denoted by $\varepsilon(1,2,3,4)$, but here we switch the notation in order to avoid any possible confusion with the dimensional regularization parameter $\epsilon$.}
\begin{equation*}
(1,2,3,4)\equiv\langle1,2\rangle[2,3]\langle3,4\rangle[4,1]-[1,2]\langle2,3\rangle[3,4]\langle4,1\rangle.
\end{equation*}

The four particle amplitude used in the computations is given by
\begin{equation}
M_4=\left(\frac{[1,2][2,3]}{\langle 3,4\rangle\langle4,1\rangle}\right)^2I_4^{1234}+\text{Permutations},
\end{equation}
where the permutations are $1243$ and $1423$. The integral $I_4^{1234}$ was already defined in Section \ref{sect:twoexamples}.

The explicit computation is presented in the next four subsections. The first one reformulates the soft theorems in terms of stripped amplitudes. The second one deals with the explicit soft limit expansion of the five particle amplitude. The third contains the computation of the action of the operators on the four particle amplitude. And the last subsection explains the result of the comparison.

\subsection{Setting up the Computation}\label{sect:example}

In order to study the soft limit relation \eqref{result0} and \eqref{result1},
%
%
it is convenient to find an equivalent statement for stripped amplitudes $M_5$ and $M_4$. As explained in \cite{Cachazo:2014fwa}, stripped amplitudes are in general defined off of the momentum-conserving support. This means that some proper definition must be given off of this support. In fact, this is the fundamental reason soft theorems are defined for full amplitudes. The prescription given in \cite{Cachazo:2014fwa} is tailored for this purpose: solve for momentum conservation using two spinors, say $\tilde\lambda_3$ and $\tilde\lambda_4$, in terms of the other data and evaluate any representation of $M_4$ and $M_5$ in such data. More explicitly, one has
\be\label{subs}
\tilde\lambda_3 = - \sum_{a\neq 3,4}^m \frac{\langle 4, a\rangle}{\langle 4, 3\rangle}\tilde\lambda_a, \quad\quad \tilde\lambda_4 = - \sum_{a\neq 3,4}^m \frac{\langle 3, a\rangle}{\langle 3, 4\rangle} \tilde\lambda_a.
\ee
For $M_5$, once the soft limit deformation parameter $\soft$ is introduced via $\lambda_5\to \soft\lambda_5$ both spinors become $\soft$-dependent
\be\label{subs2}
\tilde\lambda_3(\soft ) = - \sum_{a=1}^2 \frac{\langle 4, a\rangle}{\langle 4, 3\rangle}\tilde\lambda_a -\soft\frac{\langle 4, 5\rangle}{\langle 4, 3\rangle}\tilde\lambda_5, \quad\quad \tilde\lambda_4(\soft ) = - \sum_{a=1}^2 \frac{\langle 3, a\rangle}{\langle 3, 4\rangle} \tilde\lambda_a-\soft\frac{\langle 3, 5\rangle}{\langle 3, 4\rangle} \tilde\lambda_5.
\ee

The soft limit relation then becomes
\begin{equation}\label{result0stripped}
\underset{\soft\to 0}{\lim} \left( \soft^3 M_{5}^{(34),D=4-2\epsilon}(\soft\lambda_5)\right) = S^{(0)}M_4^{(34),D=4-2\epsilon}
\end{equation}
and
\begin{equation}\label{result1stripped}
\underset{\soft\to 0}{\lim} \left( \soft^2 M_{5}^{(34),D=4-2\epsilon}(\soft\lambda_5 ) - \frac{1}{\soft} S^{(0)}M_4^{(34),D=4-2\epsilon }\right) = S^{(1)}M_4^{(34),D=4-2\epsilon}
\end{equation}
%
%
%
where the superscript $(34)$ indicates the two spinors that were selected.

In order to study the soft theorems we have to find a Laurent expansion of $M_{5}^{(34),D=4-2\epsilon}(\soft\lambda_s )$ around $\soft =0$ before performing any expansions in $\epsilon$. More explicitly, we compute
\be\label{eq:5ptlaurent}
M_{5}^{(34),D=4-2\epsilon}(\soft\lambda_s )=\frac{C_{-3}(\epsilon)}{\soft^3}+\frac{C_{-2}(\epsilon)}{\soft^2}+\mathcal{O}(\soft^{-1}),
\ee
where $C_{-3}(\epsilon),C_{-2}(\epsilon)$ are given in terms of linear combinations of scalar integrals.

Note that since $M_{5}^{(34),D=4-2\epsilon}$ behaves as $\soft^{-3}$ as $\soft\to0$, the l.h.s.~of \eqref{result0stripped} is identical to $C_{-3}(\epsilon)$. Once we manage to confirm the validity of \eqref{result0stripped}, then the l.h.s.~of \eqref{result1stripped} becomes identical to $C_{-2}(\epsilon)$.

Of course, in practice $C_{-3}(\epsilon),C_{-2}(\epsilon)$ themselves are complicated functions of the unconstrained kinematics data (i.e., without $\tilde\lambda_3,\tilde\lambda_4$) and $\epsilon$, and so in the check we further expand $C_{-3}(\epsilon),C_{-2}(\epsilon)$ around $\epsilon=0$ and perform the comparison at each specific order in $\epsilon$. We suspect that there could be a way of carrying out the comparison at the integrand level, i.e., without performing the $\epsilon$ expansion. It would be very interesting to explore this further.

For later convenience let us define
\begin{equation*}
k'_3:=- \lambda_3\sum_{a=1}^2 \frac{\langle 4, a\rangle}{\langle 4, 3\rangle}\tilde\lambda_a, \quad
k'_4:=- \lambda_4\sum_{a=1}^2 \frac{\langle 3, a\rangle}{\langle 3, 4\rangle} \tilde\lambda_a, \qquad
p_3:=-\frac{\langle 4, 5\rangle}{\langle 4, 3\rangle}\lambda_3\tilde\lambda_5, \quad
p_4:=-\frac{\langle 3, 5\rangle}{\langle 3, 4\rangle} \lambda_4\tilde\lambda_5.
\end{equation*}
Note that with these definitions, $k_1+k_2+k'_3+k'_4=0$ has the form of $4$-point momentum conservation constraint, and so it is natural to also introduce $s'=(k_1+k_2)^2$, $u'=(k_1+k'_3)^2$, and $t'=(k_1+k'_4)^2$.

\subsection{Explicit Soft Limit Expansion of $M_5^{(34),D=4-2\epsilon}$}\label{sect:5ptside}

In this subsection we perform the $\soft$-expansion of $M_{5}^{(34),D=4-2\epsilon}(\soft\lambda_5 )$
. As discussed in previous sections, it is crucial that the soft limit $\soft$-expansion be performed before doing the $\epsilon$-expansion of the loop integrals. In practice, this means that the $\soft$-expansion is performed at the integrand level and the result is then integrated.

In order to determine the order in $\soft$ to which the integrand of the scalar integrals must be expanded, it is important to understand the $\soft$ behavior of their corresponding coefficients in $M_{5}^{(34),D=4-2\epsilon}(\soft\lambda_5 )$ as given in \eqref{eq:5ptformula}.

\begin{figure}[tbp]
	\centering
		\includegraphics[width=0.8\textwidth]{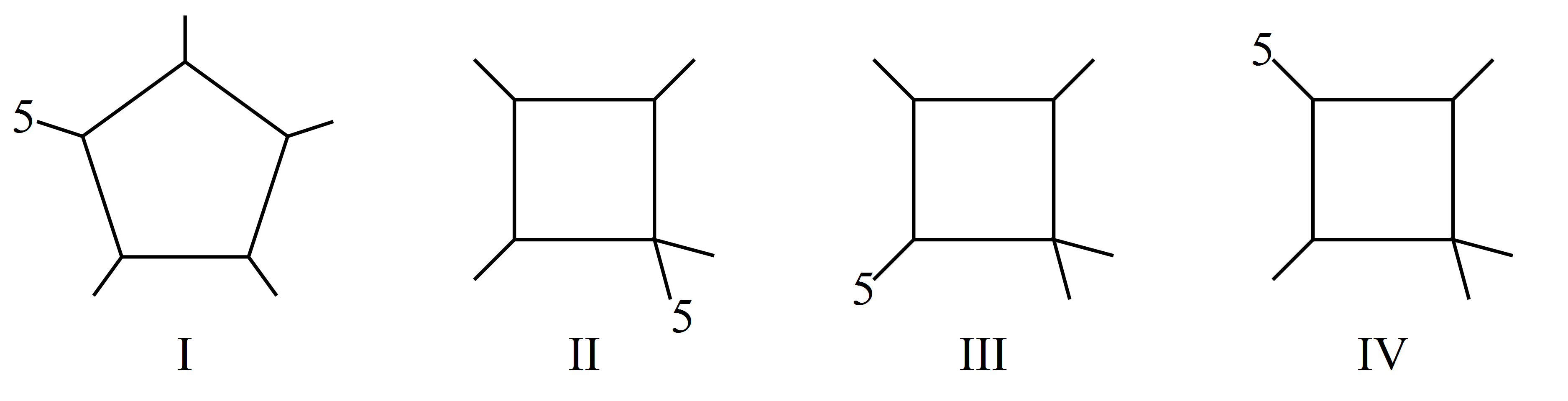}
	\caption{Pentagon and box scalar integrals involved in the soft limit expansion.}
	\label{fig:figure1}
\end{figure}

Recall that we are only interested in terms up to order $\soft^{-2}$ in the expansion. Altogether, there are four types of terms to be considered, corresponding to the graphs shown in Figure \ref{fig:figure1}.
Simple counting shows the following leading behaviors of the different coefficients\footnote{Note that since $(1,2,3,4)$ is completely anti-symmetric, by momentum conservation $(1,2,3,4)=(5,1,2,3)$, and so it actually scales as $\soft^1$.}
\begin{center}
\begin{tabular}{c||c|c|c|c}
\hline
Integral type&I&II&III&IV\\
\hline
Leading behavior of coefficients &$\soft^{-2}$&$\soft^{-3}$&$\soft^{-2}$&$\soft^{-2}$\\
\hline
\end{tabular}
\end{center}

Since the expansion of the {\it integrands} cannot have poles in $\soft$, we see that the Laurent expansion around $\soft=0$ of $M_{5}^{(34),D=4-2\epsilon}(\soft\lambda_5 )$ starts at order $\soft^{-3}$ as expected.

From the table we see that except for scalar integrals of type II, where the soft particle $5$ is attached to the massive corner of the box, the integrand of the corresponding scalar integrals has to be expanded only to leading order. In other words, it is enough to take the strict limit $k_5\to 0$ in the scalar integrals of types I, III and IV and then compute them in an $\epsilon$-expansion. Type II integrals are more complicated. For this type, we have to expand the integrand and keep corrections to order $\soft$. The answer is then integrated and expanded in $\epsilon$.

Let us now proceed to the evaluation of the corresponding contributions.

\subsubsection{Type I, III, and IV}

In type I, we choose, without loss of generality, a specific ordering, $I_5^{12345}$. After introducing the $\soft$ deformation the integral under consideration reads
\begin{equation*}
\int\frac{d^{4-2\epsilon}L}{N}\,\frac{1}{(L\!+\!k_2\!+\!k'_3\!+\!\soft\,p_3)^2\,(L\!+\!k'_3\!+\!\soft\,p_3)^2\,L^2\,(L\!-\!k'_4\!-\!\soft\,p_4)^2\,(L\!-\!k'_4\!-\!\soft\,(p_4\!+\!k_5))^2}.
\end{equation*}

The leading term in the $\soft$-expansion of the integrand leads to the following scalar box integral
\begin{equation}\label{eq:leadingtermexample}
\int\frac{d^{4-2\epsilon}L}{N}\,\frac{1}{(L+k_2+k'_3)^2\,(L+k'_3)^2\,L^2\,((L-k'_4)^2)^2},
\end{equation}
where the propagator $(L-k'_4)^2$ has weight $2$. Together with the $\soft^{-2}$ terms in the expansion of the coefficient of $I_5^{12345}$, this leading term \eqref{eq:leadingtermexample} produces the total contribution of this pentagon integral to the order $\soft^{-2}$ in the soft limit expansion of the amplitude.

Now we have to compute this integral as an expansion in $\epsilon$. Introducing the standard Feynman parameters and applying Mellin-Barnes techniques \cite{Smirnov:2004ym}, the expansion can easily be obtained as detailed in Appendix \ref{app:5ptdetail}. The $\epsilon$ expansion starts at order $\epsilon^{-2}$.

Type III and type IV terms can be analyzed in exactly the same way. For each scalar integral we are again only interested in the leading order in $\soft$, which now turns the box integrals into a scalar triangle integral with two massless corners and one of the loop propagators with weight $2$. Explicit expansion in $\epsilon$ can also be performed easily; the leading order in these cases is $\epsilon^{-1}$.

\subsubsection{Type II}

Type II integrals require more work. The reason is that their integrand has to be expanded up to order $\soft$.

As a typical example let us study
\begin{equation*}
I_4^{123(45)}=\int\frac{d^{4-2\epsilon}L}{N}\,\frac{1}{(L+k_1+k_2)^2\,(L+k_2)^2\,L^2\,(L-k'_3-\soft p_3)^2}.
\end{equation*}
In order to perform the $\soft$-expansion and then the $\epsilon$-expansion, it is best to first decompose the loop momentum $L$ in $4-2\epsilon$ dimensions as
\begin{equation*}
L=l+\mu,
\end{equation*}
where $l$ lives in the same $4$ dimensional space as the external data, and $\mu$ in its $-2\epsilon$ dimensional complement. Since $\mu$ is now orthogonal to $l$ as well as all the kinematics data, the integral can be expressed as
\begin{equation*}
I_4^{123(45)}=\!\int\frac{d^{-2\epsilon}\mu}{N}\!\int d^4l\,\frac{1}{((l\!+\!k_1\!+\!k_2)^2\!-\!\mu^2)\,((l\!+\!k_2)^2\!-\!\mu^2)\,(l^2\!-\!\mu^2)\,((l\!-\!k'_3\!-\!\soft\,p_3)^2\!-\!\mu^2)}.
\end{equation*}
This time in the $\soft$-expansion of the integrand, we need to keep track of the first order $\soft$ corrections as well. The relevant part of the expanded integrand is
\begin{equation}
\begin{split}
&\frac{1}{((l\!+\!k_1\!+\!k_2)^2\!-\!\mu^2)\,((l\!+\!k_2)^2\!-\!\mu^2)\,(l^2\!-\!\mu^2)\,((l\!-\!k'_3)^2\!-\!\mu^2)}\\
&\qquad\qquad\qquad\qquad\qquad+\frac{2\soft\,l\cdot p_3}{((l\!+\!k_1\!+\!k_2)^2\!-\!\mu^2)\,((l\!+\!k_2)^2\!-\!\mu^2)\,(l^2\!-\!\mu^2)\,((l\!-\!k'_3)^2\!-\!\mu^2)^2}
\end{split}
\end{equation}
The leading term leads to the usual scalar box integral, while the sub-leading term gives rise to a tensor integral, involving a non-trivial numerator $2\soft\,l\cdot p_3$. Since $l$ is in $4$ dimensions, we can decompose it onto a basis as follows
\begin{equation}\label{eq:ldecomposition}
l^\nu=c_1\,k_1^\nu+c_2\,k_2^\nu+c_3\,{k'_3}^\nu+c_4\,\varepsilon^{\nu\rho\sigma\omega}k_{1\rho}k_{2\sigma}k'_{3\omega}.
\end{equation}

The coefficients $c'$s can easily be computed. One first obtains three linear equations for the first three coefficients $\{c_1,c_2,c_3\}$ by contracting both sides of the equation above with $k_1$, $k_2$ and $k'_3$ respectively. Note that $c_4$ does not enter in any of these equations since $\varepsilon^{\nu\rho\sigma\omega}k_{1\rho}k_{2\sigma}k'_{3\omega}$ vanishes when contracted with any of the three $k's$.

Solving the three linear equations gives $\{c_1,c_2,c_3\}$ as functions of $\{l\cdot k_1,l\cdot k_2,l\cdot k'_3\}$ and other Lorentz invariant products formed out of the $k's$.

In order to deal with the last coefficient, $c_4$, let us define $q^\nu=\varepsilon^{\nu\rho\sigma\omega}k_{1\rho}k_{2\sigma}k'_{3\omega}$. Then we can solve for $c_4$ by contracting \eqref{eq:ldecomposition} with $q^\nu$. This gives $c_4=\frac{l\cdot q}{q^2}$.

The first observation is that when we evaluate the $l$ integral the $c_4$ term in \eqref{eq:ldecomposition} leads to the integral
\begin{equation}\label{eq:psuedointegral}
\frac{2\,p_3\cdot q}{q^2}
\int d^4l\frac{\varepsilon^{\nu\rho\sigma\omega}l_{\nu}k_{1\rho}k_{2\sigma}k'_{3\omega}}{((l+k_1+k_2)^2-\mu^2)\,((l+k_2)^2-\mu^2)\,(l^2-\mu^2)\,((l-k'_3)^2-\mu^2)^2}.
\end{equation}
As noticed in \cite{vanNeerven:1983vr}, this integral \eqref{eq:psuedointegral} is zero when the integration contour is parity invariant as it is in our case.

At this point, it is easy to find the numerator of the integrand contributing to the first order $\soft$ correction as a linear combination of $l\cdot k_1$, $l\cdot k_2$, and $l\cdot k'_3$. The coefficients are simply scalar contractions of $k's$ and $p_3$.

Applying the relations
\begin{equation*}
\begin{split}
2\,l\cdot k_1&=((l+k_1+k_2)^2-\mu^2)-((l+k_2)^2-\mu^2)-2k_1\cdot k_2,\\
2\,l\cdot k_2&=((l+k_2)^2-\mu^2)-(l^2-\mu^2),\\
2\,l\cdot k'_3&=(l^2-\mu^2)-((l-k'_3)^2-\mu^2),
\end{split}
\end{equation*}
and then recombining $l$ and $\mu$ to recover the original $L$ integration, we express the first order $\soft$ correction as a linear combination of the usual scalar integrals, with rational function coefficients.

At this order in $\soft$, the scalar integrals that appear are: (i) massless box integrals, (ii) massless box integrals with one propagator of weight $2$, and (iii) triangle integrals with two massless corners and one propagator of weight $2$. All these integrals are to be evaluated in $4-2\epsilon$ dimensions. Luckily, they have already appeared in the previous types and therefore their $\epsilon$-expansions computed. The final result is obtained again by combining with the $\soft$-expansion of the coefficient of $I_5^{123(45)}$ in the original formula.

\subsubsection{Summary of Contributions}

Here we summarize how each order in the $\soft$-expansion and $\epsilon$-expansion receives contributions from different types of terms in (\ref{eq:5ptformula}). The information is most easily presented in the form of a table
\begin{center}
\begin{tabular}{c||c|c|c}
\hline
&$\epsilon^{-2}$&$\epsilon^{-1}$&$\epsilon^{0}$\\
\hline \hline
$\soft^{-3}$&II&II&II\\
\hline
$\soft^{-2}$&I, II&I, II, III, IV&I, II, III, IV\\
\hline
\end{tabular}
\end{center}


At order $\soft^{-2}$, the fact that contributions from different types of terms combine to produce a result that exactly matches the action of $S^{(1)}$ on the $4$-point amplitude is a very nontrivial check of the new soft theorem.

Furthermore, it is well known that in gravitational amplitudes the order $\epsilon^{-2}$ should vanish \cite{Dunbar:1995ed}. It is also a non-trivial verification of the consistency of our analysis that the contributions from the type I pentagon integrals and those from the type II box integrals completely cancel each other at this order.

\subsection{Operators Acting on $M_4^{(34),D=4-2\epsilon}$}

Having completed the computation of the soft limit expansion of the $5$-point amplitude we now turn to the actions of the operators $S^{(0)}$ and $S^{(1)}$ on the $4$-point amplitude, i.e., $S^{(0)}M_4^{(34),D=4-2\epsilon}$ and $S^{(1)}M_4^{(34),D=4-2\epsilon}$.

In this section we use the following formula for the $4$-point amplitude \cite{Bern:1998sv},
\begin{equation}\label{eq:4ptformula}
M_4=\left(\frac{[1,2][2,3]}{\langle3,4\rangle\langle4,1\rangle}\right)^2I_4^{1234}+\text{Permutations},
\end{equation}
where the permutations are $1243$ and $1423$.

In obtaining $M_4^{(34),D=4-2\epsilon}$, we solve $\tilde\lambda_3,\tilde\lambda_4$ as well but this time from $4$-point momentum conservation
\begin{equation*}
\tilde\lambda_3 = - \sum_{a=1}^2 \frac{\langle 4, a\rangle}{\langle 4, 3\rangle}\tilde\lambda_a, \quad\quad \tilde\lambda_4 = - \sum_{a=1}^2 \frac{\langle 3, a\rangle}{\langle 3, 4\rangle} \tilde\lambda_a.
\end{equation*}
Note that here we are using the same unconstrained data as that in $M_5^{(34),D=4-2\epsilon}$. Since in $M_4^{(34),D=4-2\epsilon}$ we are not hitting any singularities of the kinematics, the operation of applying $S^{(0)}$ and $S^{(1)}$ commutes with the $\epsilon$-expansion. This means, on the $4$-point side we are justified to directly take the $\epsilon$-expanded formula for $M_4^{(34),D=4-2\epsilon}$, by using the $\epsilon$-expansion of the massless box integrals, e.g.,
\begin{equation}\label{eq:masslessboxexpansion}
\begin{split}
I_4^{1234}&=\int\frac{d^{4-2\epsilon}L}{N}\,\frac{1}{(L+k_1+k_2)^2\,(L+k_2)^2\,L^2\,(L-k'_3)^2}\\
&=\frac{2}{s't'}\epsilon^{-2}-\frac{\log(-s')+\log(-t')}{s't'}\epsilon^{-1}+\frac{-2\pi^2+3\log(-s')\log(-t')}{3s't'}+\mathcal{O}(\epsilon^1).
\end{split}
\end{equation}
Then the action of $S^{(0)}$ and $S^{(1)}$ produces the r.h.s.~of \eqref{result0stripped} and \eqref{result1stripped} respectively.

\subsection{Final Comparison and Comments on the Numerical Analysis}

Having explained in great detail what operations to perform on the $5$-point side and on the $4$-point side, the final step is to compare them and check if the results on both sides match each other.

Several parts of these computations must be done symbolically: (i) on the $4$-point side we have to take derivatives before we evaluate the $\lambda,\tilde\lambda$ variables on any specific set of numbers, (ii) on the $5$-point side we $\soft$-expand the integrand before any integration is done, and the way to perform the integration is to map it to the standard scalar integrals, whose results in terms of $\epsilon$-expansions are already known from literature.

The final expression obtained after these operations is complicated and in order to perform the comparison it is best to use numerical data in the end.

To be specific, the data that we need to specify in $M_5^{(34),D=4-2\epsilon}(\soft\lambda_5)$, $S^{(0)}M_4^{(34),D=4-2\epsilon}$ as well as $S^{(1)}M_4^{(34),D=4-2\epsilon}$ are
\be\label{numr}
\{\lambda_1,\lambda_2,\lambda_3,\lambda_4,\lambda_5\},\quad \{ \tilde\lambda_1,\tilde\lambda_2,\tilde\lambda_5\},
\ee
and they can be evaluated on any set of numbers, since there is no longer any constraint from momentum conservation.

With the help of some simple number theory, we can still perform the check without having to actually rely on numerical techniques. In other words, whatever conclusion of the comparison we make is \emph{exact}. The idea is to choose the data (\ref{numr}) to be rational numbers. The results will then separate according to transcendentality and will not mix, e.g., the coefficient of $\pi^2$ cannot mix with that of a logarithm of a rational number.
\begin{figure}[tbp]
	\centering
		\includegraphics[width=0.6\textwidth]{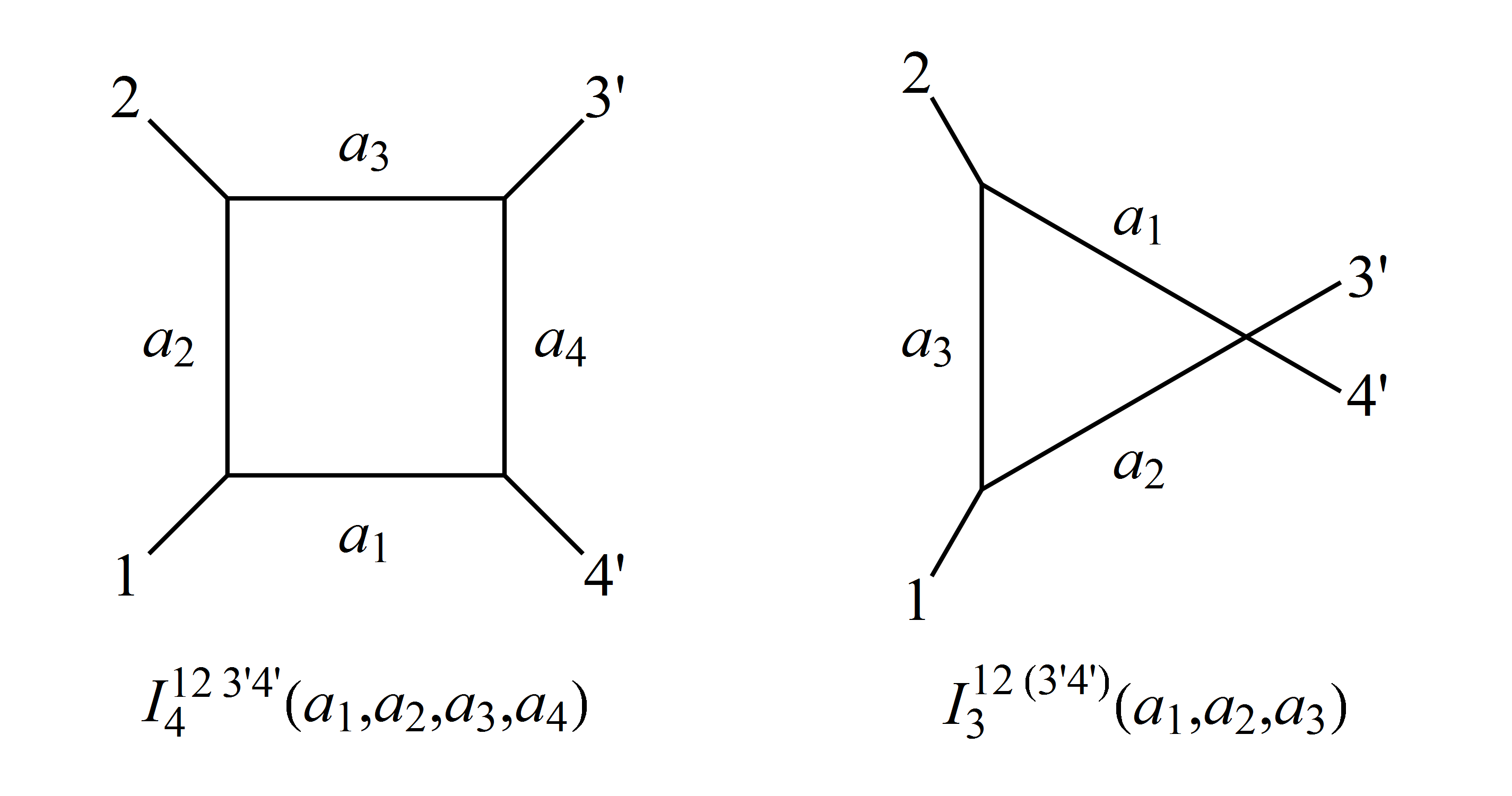}
	\caption{Examples of the scalar integrals needed in the check. The $a$'s denote the weights of the corresponding propagators. The label on each external legs denotes the momentum flowing out of that leg.}
	\label{fig:figure2}
\end{figure}

Let us start by reviewing the several scalar integrals needed for this check together with their $\epsilon$-expansions. Firstly there is the usual massless box integral, already shown in \eqref{eq:masslessboxexpansion}. As is well known, such an integral has uniform transcendentality order by order in its $\epsilon$-expansion. Then, as shown in Figure \ref{fig:figure2}, we have again the massless box integral, but with one of the propagators with weight $2$ (i.e., it is squared). For example,
\begin{align}
&I_{4}^{123'4'}(2,1,1,1)=\frac{2}{s'^2t'}\epsilon^{-2}+\frac{s'+4t'-t'\,(\log(-s')+\log(-t'))}{s'^2t'^2}\epsilon^{-1}+\\
&\quad+\frac{3s'-2\pi^2\,t'-3s'\,\log(-t')-6t'\,(\log(-s')+\log(-t'))+3t'\,\log(-s')\log(-t')}{3s'^2t'^2}+\mathcal{O}(\epsilon^1).\nonumber
\end{align}
For such weighted box integral, transcendentality levels mix with the $\epsilon$ expansion. Finally we have the two types of triangle integrals
\begin{align}
I_{3}^{12(3'4')}(2,1,1)&=\frac{1}{s'^2}\epsilon^{-1}-\frac{\log(-s')}{s'^2}+\mathcal{O}(\epsilon),\\
I_{3}^{12(3'4')}(1,1,2)&=-\frac{1}{s'^2}\epsilon^{-1}+\frac{1+\log(-s')}{s'^2}+\mathcal{O}(\epsilon).
\end{align}
%


Recalling that all external kinematics are rational numbers, at each specific $\soft$ and $\epsilon$ order, we are able to keep track of the purely rational terms, terms proportional to $\pi^2$, terms proportional to $\log$, and terms proportional to $\log\,\log$. In fact, it is not hard to observe that there are only six types of terms that depend on $\log$, which are: $\log(-s')$, $\log(-t')$, $\log(-u')$, $\log(-s')\log(-t')$, $\log(-u')\log(-t')$, and $\log(-s')\log(-u')$ (with $\{s',t',u'\}$ defined in the end of Subsection \ref{sect:example}).

Although in the formulas above we observe mixing of different transcendentalities in a given order of $\epsilon$, after we compute the result of the full amplitude, (on both sides) each specific order of $\soft$ and $\epsilon$ again recovers uniform transcendentality. In particular, at the order $\epsilon^{-1}$ all the rational terms cancel away and we are left with only terms proportional to $\log$.

The conclusion is that our analysis shows that the soft theorems
\begin{equation}
\underset{\soft\to 0}{\lim} \left( \soft^3 M_{5}^{(34),D=4-2\epsilon}(\soft\lambda_s)\right) = S^{(0)}M_4^{(34),D=4-2\epsilon}.
\end{equation}
and
\begin{equation}
\underset{\soft\to 0}{\lim} \left( \soft^2 M_{5}^{(34),D=4-2\epsilon}(\soft\lambda_s ) - \frac{1}{\soft} S^{(0)}M_4^{(34),D=4-2\epsilon }\right) = S^{(1)}M_4^{(34),D=4-2\epsilon}.
\end{equation}
hold exactly at orders $\epsilon^{-2}$, $\epsilon^{-1}$, and $\epsilon^0$. It would be important to check these relations to higher orders in $\epsilon$.



\section{Box Expansion vs.~All Orders in $\epsilon$ Formulas}\label{sect:correctformula}

In this section, we provide further detailed evidence of why a formula valid to all orders in $\epsilon$ is needed when studying soft limits.

To illustrate this in detail, let us go back to the $5$-point side in our analysis. Since we are checking the new soft theorems only up to order $\epsilon^0$ in dimensional regularization, one would be tempted to start with what is known as a box expansion. In
\cite{Bern:1998sv} it was shown that
\begin{equation}\label{eq:5ptformulabox}
M_5=\frac{1}{2}\sum\frac{[1,2]^2[2,3]^2[4,5]}{\langle4,5\rangle\langle3,4\rangle\langle3,5\rangle\langle4,1\rangle\langle5,1\rangle}
I_{4}^{123(45)}+\mathcal{O}(\epsilon^1).
\end{equation}
Note that here $I_{4}^{123(45)}$ denotes a scalar integral before any $\epsilon$ expansion.

As we will see, this formula does not reproduce the soft theorems correctly. Here the summation is again over all the $30$ inequivalent one-mass box integrals. In order to see what is going wrong, let us use this formula literally, ignoring the $\mathcal{O}(\epsilon^1)$ corrections. Following the same analysis described in the previous sections, we find that in the soft limit one gets the same result at all orders we checked in Section 3 except for the contribution at order $\soft^{-2}\epsilon^0$. The difference only involves terms proportional to logarithms, which have the wrong transcendentality regarding this power of $\epsilon$. This is an indication that they have to be canceled away by something different from regular boxes.

To explain the origin of this problem, let us first recall that the representation \eqref{eq:5ptformulabox} can be derived from the fully valid formula \eqref{eq:5ptformula} by doing van Neerven--Vermaseren reduction to the scalar pentagon integrals therein to decompose them into scalar boxes \cite{vanNeerven:1983vr}. Note that the pentagon integrals in \eqref{eq:5ptformula} are evaluated in $4-2\epsilon$ dimensions. The decomposition of such a pentagon integral into box integrals reads \cite{Bern:2006vw}
\begin{equation}\label{eq:pentagondecomposition}
I^{12345}_{5,D=4-2\epsilon}=\sum_{\text{cyclic}} F^{123(45)}\,I^{123(45)}_{4,D=4-2\epsilon}-\epsilon\,\frac{(1,2,3,4)^2}{s_{1,2}s_{2,3}s_{3,4}s_{4,5}s_{5,1}}\,I^{12345}_{5,D=6-2\epsilon},
\end{equation}
where the rational functions $F^{123(45)}=F^{123(45)}(\{k\},\soft)$ are finite to the leading order in $\soft$, which is irrelevant in the current discussion. In the above we have explicitly shown the dimensions in which the integrals should be evaluated.

Now this is when the second example presented in Subsection \ref{sect:twoexamples} becomes useful. As mentioned in Subsection \ref{sect:twoexamples}, scalar pentagon integrals are $\mathcal{O}(\epsilon^0)$ in $6-2\epsilon$ dimensions, and thus we see that this last term in \eqref{eq:pentagondecomposition} is the source of the $\mathcal{O}(\epsilon^1)$ corrections in \eqref{eq:5ptformulabox}. Explicitly, the missing part in \eqref{eq:5ptformulabox} is
\begin{equation}
-\epsilon\,\frac{[1,2][2,3][3,4][4,5][5,1]}{\langle1,2\rangle\langle2,3\rangle\langle3,4\rangle\langle4,5\rangle\langle5,1\rangle}\,I^{12345}_{5,D=6-2\epsilon}+\text{Permutations},
\end{equation}
where the permutations are again over the $12$ inequivalent pentagon integrals. The striking thing is that, although for generic kinematics data $I_{5,D=6-2\epsilon}$ remains finite as $\epsilon\to0$, after the soft limit it diverges as $\epsilon^{-2}$. And so these ``irrelevant'' terms start to impose real effects in four dimensions. In the end, each term gives non-zero contributions starting at order $\soft^{-2}$ and order $\epsilon^{-1}$. Quite nicely, the summation over the $\epsilon^{-1}$ terms completely gives zero  and hence it does not affect the result from the boxes alone. When we take the summation of the $\epsilon^{0}$ terms into consideration, we exactly fix the problem encountered at order $\soft^{-2}\epsilon^0$ from the use of only boxes.

\section{Soft Theorems vs.~Soft Expansions}

The point of view we have advocated in this work is that there are two ways of treating relations among S-matrix elements. One can consider a given matrix element for the scattering of $n$ particles and then expand its stripped amplitude when say $0 < s_{an}/s_{ab}\ll 1$ for all $a$ and $b$ in $\{1, \ldots ,n-1\}$. This is what we would like to call a soft expansion. Since this is meant to describe the behavior of an $n$-particle amplitude in a certain region of the space of kinematic invariants, it is natural to first perform any loop integrals over the full range of the available phase space as an expansion in the regulator parameter. This soft expansion is very physical and useful, e.g., in QCD computations the discontinuities that arise are important in the cancellation of real emission divergences.

The second point of view is not an expansion of a given amplitude but relations among distributions constructed from S-matrix elements. As mentioned in the introduction, this is motivated by the study of the relations among S-matrix elements derived from the action of the BMS group and its extensions \cite{Strominger:2013un,Strominger:2013jfa,He:2014laa}. The corresponding S-matrix elements can have different number of particles. These relations hold in the strict limit $s_{an}/s_{ab} \to 0$. These are what we would like to call soft theorems. At tree-level the precise formulation of soft theorems is not relevant but at loop level a number of subtleties arise. These come from two main sources. the first is that in general, scattering amplitudes of momentum eigenstates are divergent and need a regulator. The second is that loop integrations explore all regions of the available phase space. To make the latter point clear consider the box one loop integral discussed in section 2.2 but now in four dimensional Minkowski space and expanded by using the Feynman tree theorem \cite{Feynman:1963ax}
\begin{equation}\label{ftt}
\begin{split}
&\int d^4{L}\,\frac{1}{(L^2+i\varepsilon)((L+k_1)^2+i\varepsilon)((L+k_1+k_2)^2+i\varepsilon)((L-k_3)^2+i\varepsilon)} =\\
&\qquad\qquad\int d^4{L}\,\delta^{(+)}(L^2)\,\frac{1}{((L+k_1)^2+i\varepsilon)((L+k_1+k_2)^2+i\varepsilon)((L-k_3)^2+i\varepsilon)}+\cdots
\end{split}\end{equation}
The ellipses denote other terms with single cuts and terms with two or more cuts. In each term, one is supposed to integrate over the full available phase space dictated by the measure. For example, the first term, shown explicitly in (\ref{ftt}) integrates over all on-shell future directed momenta $L$. Clearly, this integration contains regions where particle $L$ is soft and regions where it is collinear with other particles or both.

Our point of view on the soft theorems, as relations among distributions which are valid for test functions with arbitrarily small support, is that they must be defined as limits which take priority over all other limits in the amplitude including the regions of $L$. More explicitly, the limit of $\tau\to 0$ must be taken before one lets the on-shell particle $L$ explore soft/collinear regions in the phase space.

The simplest way to implement the prescription is to perform all expansions for computing limits in $\tau$ at the integrand level and then perform the integrations while keeping the regulator finite. Only when $\tau$ does not enter in the resulting expression can other expansions or limits be done.


\section{Discussions and Future Directions}\label{sect:discussion}

In this work we have shown that relations among S-matrix elements coming from soft limits have to be defined as relations among distributions. This in turn does not have much effect at tree level but it leads to dramatic consequences at loop level. The most important of these is that expressions for scattering amplitudes must be kept in integral form while the soft limit expansion is performed. More explicitly, the soft limit expansion is performed at the level of the integrand. It is only when this process is completed that integrations can be performed and expanded in the dimensional regularization parameter $\epsilon$.

We illustrated the distributional perspective with soft theorems applied to a five-point one-loop scattering amplitude in ${\cal N}=8$ supergravity. Quite non-trivially we found that the leading and sub-leading soft theorems are \emph{not} renormalized. It would be very important to explore this point of view in other theories of gravity and in Yang-Mills (for recent discussions on gauge theory see \cite{Casali:2014xpa,Larkoski:2014hta}).

It is well-known that the tree-level version of Weinberg's soft theorem does not hold exactly when applied to $\epsilon$-expanded Yang-Mills amplitudes. It is very natural to suggest that if the distributional point of view is taken then the loop corrections to the soft theorem can disappear. Moreover, in theories with color ordering it is possible to defined a physical integrand to all loop orders \cite{ArkaniHamed:2010kv,Boels:2010nw} in the large $N$ limit and therefore one would expect the soft theorems to hold already at the integrand level.

In this work, we only performed the analysis of the first two soft theorems for gravity coming from the leading and sub-leading terms. In \cite{Cachazo:2014fwa}, a third theorem was shown to hold at tree-level using a second order differential operator called $S^{(2)}$. The operator $S^{(2)}$ controls the order $\soft^{-1}$ in the expansion. It would be fascinating to extend the analysis to this order in $\soft$.

As mentioned above, it is important to explore loop amplitudes in pure gravity, i.e., without supersymmetry. Two series of amplitudes stand out in pure gravity. These are the all-plus, ${\cal M}(1^+,2^+,3^+,\ldots , n^+)$, and the one-minus, ${\cal M}(1^-,2^+,3^+,\ldots , n^+)$. The reason these are special is that they are purely rational functions of the kinematic invariants and are free of UV and IR divergences. Our results show that in order to study soft limits for these amplitudes one actually has to find a representation for them in terms of one-loop integrals before any expansion in $\epsilon$ is done. The reason is that, as we have seem in different forms in this work, integrals that normally contribute to order $\epsilon^1$ can become of order $\epsilon^0$ when the soft limit is applied before expanding in $\epsilon$.

In \cite{He:2014bga}, the application of the soft theorem involving $\soft^{-3}S^{(0)}$, $\soft^{-2}S^{(1)}$ and $\soft^{-1}S^{(2)}$ directly to the $\epsilon$-expanded all-plus and one-minus amplitudes was studied. It was verified up to twelve particles that all three orders in $\soft$ correctly reproduce the soft limit in the all-plus sector and that only the first two orders coincide with the expansion for the one-minus sector. We would like to interpret these results as good indication that if the computation is done using the order of limits implied by our work then all three orders of the theorems will exactly hold.

It is also natural to suspect that all relations among scattering amplitudes of momentum eigenstates, such as collinear and factorization limits must also be interpreted in the distributional sense. Unitarity then becomes a statement about the strict limit. Consider for example a multi-particle factorization where the sum of some subset of external data becomes null, say $k_1+k_2+\cdots +k_m$.  In this particular case one has $(k_1+k_2+\cdots +k_m)^2 \to 0$ and therefore unitarity should be expressed as
\be
\underset{(k_1+k_2+\cdots +k_m)^2\to 0}{\lim}{\cal M}_{k_1,k_2,\ldots , k_n} = \int d^4q \delta(q^2){\cal M}_{k_1,\ldots ,k_m,q}{\cal M}_{-q,k_{m+1},\ldots ,k_n}.
\ee
Just as in the discussion of the soft limits, once dealing with an interacting theory a choice of regulator must be made. Choosing again to work in $4-2\epsilon$ dimensions we learn that relations among S-matrix elements must be interpreted in such a way that the kinematic limit is taken first and only after the expansion of integrands is completed then the integration and corresponding expansion in $\epsilon$ must be performed.

Finally, it is important to mention that in this paper we have assumed the standard definition of the S-matrix for momentum eigenstates. However, in theories of quantum gravity such a definition might not be the most physical one. Recent work \cite{Ware:2013zja} shows that using techniques adapted from QED it is possible to construct an IR finite S-matrix for gravity. It would be very interesting to find out how the soft theorems discussed here translate into their construction.

\acknowledgments

FC would like to thank P.~Schuster, N.~Toro and I.~Yavin for discussions. FC and EYY are both grateful to A.~Strominger for useful discussions and comments on the manuscript. This work is supported by the Perimeter Institute for Theoretical Physics. Research at Perimeter Institute is supported by the Government of Canada through Industry Canada and by the Province of Ontario through the Ministry of Research \& Innovation.


\appendix


\section{Details about the Integrals Needed in the Five-Point Analysis}\label{app:5ptdetail}

In the analysis of $\soft$-expansion of $5$-point amplitude in Subsection \ref{sect:5ptside}, we have encountered several types of scalar integrals, whose result in terms of $\epsilon$-expansion is in need for the purpose of comparison with the $4$-point side. In this appendix we collect necessary facts about these integrals.

To be specific, the integrals appearing in the analysis are (i) massless scalar box integrals, possibly with one of the loop propagators of weight $2$, and (ii) scalar triangle integrals with two massless corners and a massive corner, and one of the loop propagators of weight $2$. Let us start form a more general definition of these integrals, where arbitrary weights are assigned to the propagators, and the loop integration is performed in $D$ dimensions.
\begin{align}
I_{4,D}^{1234}(a_1,a_2,a_3,a_4)&=\int\frac{d^{D}L}{N}\,\frac{1}{((L+k_1+k_2)^2)^{a_1}((L+k_2)^2)^{a_2}(L^2)^{a_3}((L-k_3)^2)^{a_4}},\\
I_{3,D}^{12P}(a_1,a_2,a_3)&=\int\frac{d^{D}L}{N}\,\frac{1}{((L+k_1)^2)^{a_1}((L-k_2)^2)^{a_2}(L^2)^{a_3}},
\end{align}
where $\{a_i\}$ are the weights associated to the corresponding propagators, and we insert the normalization factor $N=2i\pi^{\frac{D}{2}}e^{-\gamma_E\epsilon}$ in order to make the $\epsilon$-expansion simpler. As before, the subscripts denote the number of propagators as well as the dimensions in which the integration is evaluated. And the superscripts denote the momenta flowing out of each corner. In particular, in the case of triangle integral we assume that the corners $1$ and $2$ are massless, while $P$ is massive. We denote $s=(k_1+k_2)^2$ and $t=(k_2+k_3)^2$. Graphs corresponding to specific examples of these integrals have been shown in Figure \ref{fig:figure2}.

By introducing Feynman parameters and applying Mellin-Barnes technique, the box integral above can be transformed to its Mellin representation
\begin{align}\label{eq:boxMellin}
I_{4,D}^{1234}(a_1,a_2,a_3,a_4)=
\int\frac{dz}{2\pi i}&\left(\frac{t}{s}\right)^z\frac{(-1)^ae^{\gamma_E\epsilon}\,\Gamma(a-\frac{D}{2}+z)\,\Gamma(a_2+z)\,\Gamma(a_4+z)\,\Gamma(-z)}{2\,\Gamma(D-a)\,\Gamma(a_1)\,\Gamma(a_2)\,\Gamma(a_3)\,\Gamma(a_4)\,(-s)^{a-\frac{D}{2}}}\cdot\nonumber\\
&\quad\cdot\Gamma(\frac{D}{2}-a_1-a_2-a_4-z)\Gamma(\frac{D}{2}-a_2-a_3-a_4-z),
\end{align}
where $a=a_1+a_2+a_3+a_4$. The integration is performed along a specially chosen contour on the complex $z$-plane from $-i\infty$ to $+i\infty$, such that it separates the poles of all Gamma functions of the form $\Gamma(A+z)$ and those of all Gamma functions of the form $\Gamma(B-z)$. In order to extract the $\epsilon$-expansion, one can either apply Barnes lemmas or perform contour deformations, depending on specific cases. Details about the technique can be found in, e.g., \cite{Smirnov:2004ym}. For our specific purpose, in the following we only list out the results that are needed in the current analysis. In $D=4-2\epsilon$ dimensions, we have
\begin{align}
\label{eq:masslessbox}
I_{4,D=4-2\epsilon}^{1234}(1&,1,1,1)=\frac{2}{st}\epsilon^{-2}-\frac{\log(-s)+\log(-t)}{st}\epsilon^{-1}+\frac{-2\pi^2+3\log(-s)\log(-t)}{3st}+\mathcal{O}(\epsilon^1).\\
\label{eq:deformedbox4d}I_{4,D=4-2\epsilon}^{1234}(2&,1,1,1)=\frac{2}{s^2t}\epsilon^{-2}+\frac{s+4t-t\,(\log(-s)+\log(-t))}{s^2t^2}\epsilon^{-1}+\\
&+\frac{3s-2\pi^2\,t-3s\,\log(-t)-6t\,(\log(-s)+\log(-t))+3t\,\log(-s)\log(-t)}{3s^2t^2}+\mathcal{O}(\epsilon^1).\nonumber
\end{align}
In the analysis in Section \ref{sect:correctformula}, we need to evaluate \eqref{eq:deformedbox4d} again but now in $D=6-2\epsilon$ instead, which is
\begin{equation}
I_{4,D=6-2\epsilon}^{1234}(2,1,1,1)=-\frac{1}{2st}\epsilon^{-2}+\frac{\log(-t)}{2st}\epsilon^{-1}+\mathcal{O}(\epsilon^0).
\end{equation}
By studying symmetries of \eqref{eq:boxMellin}, in any dimensions we always have the following relations
\begin{align}
I_{4,D}^{1234}(1,2,1,1)&=I_{4,D}^{2341}(2,1,1,1),\\
I_{4,D}^{1234}(1,1,2,1)&=I_{4,D}^{1234}(2,1,1,1),\\
I_{4,D}^{1234}(1,1,1,2)&=I_{4,D}^{1234}(1,2,1,1),
\end{align}
which can be used to obtain the other needed integrals.

By applying the same techniques, one can work out the final result of the triangle integral defined above for generic weights, which is
\begin{equation}
I_{3,D}^{123}(a_1,a_2,a_3)=\frac{e^{\gamma_E\epsilon}}{2}\frac{\Gamma(\frac{D}{2}-a_1-a_3)\,\Gamma(\frac{D}{2}-a_2-a_3)\,\Gamma(a_1+a_2+a_3-\frac{D}{2})}{\Gamma(a_1)\,\Gamma(a_2)\,\Gamma(D-a_1-a_2-a_3)\,(-s)^{a_1+a_2+a_3-\frac{D}{2}}}.
\end{equation}
In $D=4-2\epsilon$, the $\epsilon$-expansion up to finite order of the specific cases we need are
\begin{align}
I_{3,D=4-2\epsilon}^{12P}(2,1,1)&=\frac{1}{s^2}\epsilon^{-1}-\frac{\log(-s)}{s^2}+\mathcal{O}(\epsilon),\\
I_{3,D=4-2\epsilon}^{12P}(1,1,2)&=-\frac{1}{s^2}\epsilon^{-1}+\frac{1+\log(-s)}{s^2}+\mathcal{O}(\epsilon).
\end{align}
And again by symmetry we have
\begin{equation}
I_{3,D}^{12P}(1,2,1)=I_{3,D}^{12P}(2,1,1).
\end{equation}

\bibliographystyle{JHEP}
\bibliography{SoftLimitsGR}

\providecommand{\href}[2]{#2}\begingroup\raggedright\begin{thebibliography}{10}

\bibitem{Cachazo:2014fwa}
F.~Cachazo and A.~Strominger, {\it {Evidence for a New Soft Graviton Theorem}},
   \href{http://xxx.lanl.gov/abs/1404.4091}{{\tt arXiv:1404.4091}}.

\bibitem{Weinberg:1964ew}
S.~Weinberg, {\it {Photons and Gravitons in s Matrix Theory: Derivation of
  Charge Conservation and Equality of Gravitational and Inertial Mass}},  {\em
  Phys.Rev.} {\bf 135} (1964) B1049--B1056.

\bibitem{Weinberg:1965nx}
S.~Weinberg, {\it {Infrared Photons and Gravitons}},  {\em Phys.Rev.} {\bf 140}
  (1965) B516--B524.

\bibitem{Strominger:2013un}
A.~Strominger, {\it {Virasoro Notes}},  {\em unpublished} (2013).

\bibitem{Strominger:2013jfa}
A.~Strominger, {\it {On BMS Invariance of Gravitational Scattering}},
  \href{http://xxx.lanl.gov/abs/1312.2229}{{\tt arXiv:1312.2229}}.

\bibitem{He:2014laa}
T.~He, V.~Lysov, P.~Mitra, and A.~Strominger, {\it {BMS Supertranslations and
  Weinberg's Soft Graviton Theorem}},
  \href{http://xxx.lanl.gov/abs/1401.7026}{{\tt arXiv:1401.7026}}.

\bibitem{Low:1958sn}
F.~Low, {\it {Bremsstrahlung of Very Low-Energy Quanta in Elementary Particle
  Collisions}},  {\em Phys.Rev.} {\bf 110} (1958) 974--977.

\bibitem{Burnett:1967km}
T.~Burnett and N.~M. Kroll, {\it {Extension of the Low Soft Photon Theorem}},
  {\em Phys.Rev.Lett.} {\bf 20} (1968) 86.

\bibitem{Gross:1968in}
D.~J. Gross and R.~Jackiw, {\it {Low-Energy Theorem for Graviton Scattering}},
  {\em Phys.Rev.} {\bf 166} (1968) 1287--1292.

\bibitem{Jackiw:1968zza}
R.~Jackiw, {\it {Low-Energy Theorems for Massless Bosons: Photons and
  Gravitons}},  {\em Phys.Rev.} {\bf 168} (1968) 1623--1633.

\bibitem{White:2011yy}
C.~D. White, {\it {Factorization Properties of Soft Graviton Amplitudes}},
  {\em JHEP} {\bf 1105} (2011) 060,
  [\href{http://xxx.lanl.gov/abs/1103.2981}{{\tt arXiv:1103.2981}}].

\bibitem{Bern:1998sv}
Z.~Bern, L.~J. Dixon, M.~Perelstein, and J.~Rozowsky, {\it {Multi-Leg One-Loop
  Gravity Amplitudes from Gauge Theory}},  {\em Nucl.Phys.} {\bf B546} (1999)
  423--479, [\href{http://xxx.lanl.gov/abs/hep-th/9811140}{{\tt
  hep-th/9811140}}].

\bibitem{He:2014bga}
S.~He, Y.-t. Huang, and C.~Wen, {\it {Loop Corrections to Soft Theorems in
  Gauge Theories and Gravity}},  \href{http://xxx.lanl.gov/abs/1405.1410}{{\tt
  arXiv:1405.1410}}.

\bibitem{Bern:2014oka}
Z.~Bern, S.~Davies, and J.~Nohle, {\it {On Loop Corrections to Subleading Soft
  Behavior of Gluons and Gravitons}},
  \href{http://xxx.lanl.gov/abs/1405.1015}{{\tt arXiv:1405.1015}}.

\bibitem{gelfand}
I.~Gel'fand and G.~Shilov, {\em Generalized Functions, Vol.1: Properties and
  Operations}.
\newblock Academic Press, 1964.

\bibitem{Bern:1991aq}
Z.~Bern and D.~A. Kosower, {\it {The Computation of Loop Amplitudes in Gauge
  Theories}},  {\em Nucl.Phys.} {\bf B379} (1992) 451--561.

\bibitem{Bern:1995ix}
Z.~Bern and G.~Chalmers, {\it {Factorization in one loop gauge theory}},  {\em
  Nucl.Phys.} {\bf B447} (1995) 465--518,
  [\href{http://xxx.lanl.gov/abs/hep-ph/9503236}{{\tt hep-ph/9503236}}].

\bibitem{Kniehl:2010aj}
B.~A. Kniehl and O.~V. Tarasov, {\it {Analytic Result for the One-Loop Scalar
  Pentagon Integral with Massless Propagators}},  {\em Nucl.Phys.} {\bf B833}
  (2010) 298--319, [\href{http://xxx.lanl.gov/abs/1001.3848}{{\tt
  arXiv:1001.3848}}].

\bibitem{Cachazo:2008vp}
F.~Cachazo, {\it {Sharpening The Leading Singularity}},
  \href{http://xxx.lanl.gov/abs/0803.1988}{{\tt arXiv:0803.1988}}.

\bibitem{Carrasco:2011mn}
J.~J. Carrasco and H.~Johansson, {\it {Five-Point Amplitudes in $N=4$
  Super-Yang-Mills Theory and $N=8$ Supergravity}},  {\em Phys.Rev.} {\bf D85}
  (2012) 025006, [\href{http://xxx.lanl.gov/abs/1106.4711}{{\tt
  arXiv:1106.4711}}].

\bibitem{Yuan:2012rg}
E.~Y. Yuan, {\it {Virtual Color-Kinematics Duality: 6-pt 1-Loop MHV
  Amplitudes}},  {\em JHEP} {\bf 1305} (2013) 070,
  [\href{http://xxx.lanl.gov/abs/1210.1816}{{\tt arXiv:1210.1816}}].

\bibitem{Smirnov:2004ym}
V.~A. Smirnov, {\it {Evaluating Feynman Integrals}},  {\em Springer Tracts
  Mod.Phys.} {\bf 211} (2004) 1--244.

\bibitem{vanNeerven:1983vr}
W.~van Neerven and J.~Vermaseren, {\it {Large Loop Integrals}},  {\em
  Phys.Lett.} {\bf B137} (1984) 241.

\bibitem{Dunbar:1995ed}
D.~C. Dunbar and P.~S. Norridge, {\it {Infinities within Graviton Scattering
  Amplitudes}},  {\em Class.Quant.Grav.} {\bf 14} (1997) 351--365,
  [\href{http://xxx.lanl.gov/abs/hep-th/9512084}{{\tt hep-th/9512084}}].

\bibitem{Bern:2006vw}
Z.~Bern, M.~Czakon, D.~Kosower, R.~Roiban, and V.~Smirnov, {\it {Two-Loop
  Iteration of Five-Point $N=4$ Super-Yang-Mills Amplitudes}},  {\em
  Phys.Rev.Lett.} {\bf 97} (2006) 181601,
  [\href{http://xxx.lanl.gov/abs/hep-th/0604074}{{\tt hep-th/0604074}}].

\bibitem{Feynman:1963ax}
R.~Feynman, {\it {Quantum theory of gravitation}},  {\em Acta Phys.Polon.} {\bf
  24} (1963) 697--722.

\bibitem{Casali:2014xpa}
E.~Casali, {\it {Soft Sub-Leading Divergences in Yang-Mills Amplitudes}},
  \href{http://xxx.lanl.gov/abs/1404.5551}{{\tt arXiv:1404.5551}}.

\bibitem{Larkoski:2014hta}
A.~J. Larkoski, {\it {Conformal Invariance of the Subleading Soft Theorem in
  Gauge Theory}},  \href{http://xxx.lanl.gov/abs/1405.2346}{{\tt
  arXiv:1405.2346}}.

\bibitem{ArkaniHamed:2010kv}
N.~Arkani-Hamed, J.~L. Bourjaily, F.~Cachazo, S.~Caron-Huot, and J.~Trnka, {\it
  {The All-Loop Integrand For Scattering Amplitudes in Planar $N=4$ SYM}},
  {\em JHEP} {\bf 1101} (2011) 041,
  [\href{http://xxx.lanl.gov/abs/1008.2958}{{\tt arXiv:1008.2958}}].

\bibitem{Boels:2010nw}
R.~H. Boels, {\it {On BCFW Shifts of Integrands and Integrals}},  {\em JHEP}
  {\bf 1011} (2010) 113, [\href{http://xxx.lanl.gov/abs/1008.3101}{{\tt
  arXiv:1008.3101}}].

\bibitem{Ware:2013zja}
J.~Ware, R.~Saotome, and R.~Akhoury, {\it {Construction of an Asymptotic S
  Matrix for Perturbative Quantum Gravity}},  {\em JHEP} {\bf 1310} (2013) 159,
  [\href{http://xxx.lanl.gov/abs/1308.6285}{{\tt arXiv:1308.6285}}].

\end{thebibliography}\endgroup
\end{document}